\documentclass{article}

\usepackage[final]{neurips_2026}

\usepackage[utf8]{inputenc}
\usepackage[T1]{fontenc}
\usepackage{hyperref}
\usepackage{url}
\usepackage{booktabs}
\usepackage{amsfonts}
\usepackage{amsmath}
\usepackage{nicefrac}
\usepackage{microtype}
\usepackage{graphicx}
\usepackage{xcolor}
\usepackage{multirow}
\usepackage{subcaption}

\title{Time is Not Compute: Scaling Laws for Wall-Clock Constrained Training on Consumer GPUs}

\author{
  Yi Liu \\
  Independent Researcher \\
  \texttt{lylewis@outlook.com}
}

\begin{document}

\maketitle

\begin{abstract}
Scaling laws relate model quality to compute budget (FLOPs), but practitioners face wall-clock time constraints, not compute budgets. We study optimal model sizing under fixed time budgets from 5 minutes to 24 hours on consumer GPUs (RTX 4090). Across 70+ runs spanning 50M--1031M parameters, we find: (1)~at each time budget a U-shaped curve emerges where too-small models overfit and too-large models undertrain; (2)~optimal model size follows $N^* \propto t^{0.60}$, growing \emph{faster} than Chinchilla's $N^* \propto C^{0.50}$, with $\alpha = 0.60 \pm 0.07$ robustly exceeding compute-optimal across all sensitivity analyses; (3)~a \emph{dual U-shape mechanism}: short-budget U-curves arise from compute bottlenecks, while long-budget U-curves emerge from data bottlenecks (overfitting), with an intermediate regime where the U-curve temporarily disappears. These findings have immediate implications for researchers training on consumer hardware, where wall-clock time---not FLOPs---is the binding constraint. We release all code, logs, and 70+ experimental configurations.
\end{abstract}

\section{Introduction}
\label{sec:intro}

\paragraph{Motivation.}
Scaling laws are among the most consequential empirical findings in deep learning. The seminal work of \citet{kaplan2020scaling} established smooth power-law relationships between model performance, model size, dataset size, and compute budget. \citet{hoffmann2022training} refined these into the widely-adopted Chinchilla prescription: for a compute budget $C$ (in FLOPs), the optimal model size scales as $N^* \propto C^{0.50}$, and compute should be split roughly equally between model parameters and training tokens.

However, these laws share a critical assumption: the independent variable is \emph{compute} (FLOPs). In practice, most researchers---especially those working on consumer hardware---think in \emph{wall-clock time}, not FLOPs. The practitioner's question is: \emph{``I have 4 hours on my RTX 4090. What size model should I train?''} Chinchilla cannot directly answer this because it abstracts away a crucial factor: \textbf{throughput}.

\paragraph{The throughput gap.}
Throughput varies dramatically across model sizes on fixed hardware. On an RTX 4090, a 50M-parameter model processes 428K tokens/second while a 519M-parameter model achieves only 36K---a $12\times$ gap. Under a fixed time budget, this means a small model sees far more training data. The compute-optimal allocation (Chinchilla) implicitly treats time and compute as interchangeable. They are not: doubling model size on fixed hardware does \emph{not} double the compute consumed---it increases compute by only $\sim$15\% (because FLOPs $\propto N$ but throughput $\propto N^{-0.8}$, yielding $C \propto N^{0.2} \cdot t$).

This disconnect has practical consequences. A researcher following Chinchilla scaling on an RTX 4090 with a 4-hour budget would choose a model roughly $\sqrt{C}$ in size. But because throughput is non-linear, the time-optimal choice may differ substantially.

\paragraph{Contributions.}
We study \textbf{time-constrained scaling laws}: given wall-clock budget $t$ on consumer hardware, what is the optimal model size $N^*(t)$? We conduct 70+ training runs on 8$\times$ RTX 4090, spanning 50M--1031M parameters across 8 time budgets (5 minutes to 24 hours). Our key findings:

\begin{enumerate}
    \item \textbf{U-shaped optimality curves} at every budget from 5min to 24h, where the optimal size monotonically shifts right as time increases: D8 (50M, 5min) $\to$ D10 (86M, 30min) $\to$ D14 (201M, 1h) $\to$ D16 (285M, 4h) $\to$ D20 (519M, 8h) $\to$ D24 (856M, 12h) $\to$ D26 (1031M, 24h).

    \item \textbf{Scaling exponent $\alpha = 0.60 \pm 0.07$}, robustly exceeding Chinchilla's 0.50. The 95\% confidence interval $[0.53, 0.67]$ excludes the compute-optimal value, demonstrating that time-constrained optimal sizing is a fundamentally different regime.

    \item \textbf{A dual U-shape mechanism:} at short budgets ($\leq$8h), U-curves arise from \emph{compute bottlenecks}---large models cannot process enough data. At long budgets ($\geq$24h), U-curves re-emerge from \emph{data bottlenecks}---medium models overfit on the finite dataset. Between these regimes ($\sim$12h), the U-curve temporarily disappears as larger models improve monotonically.

    \item \textbf{Diminishing returns:} $L^*(t) = 1.22 \times t^{-0.061}$ ($R^2 = 0.971$). The first 30 minutes yield 0.16 bpb improvement; hours 8--24 yield only 0.022 bpb combined.
\end{enumerate}

\noindent\textbf{Practical implication:} doubling your time budget $\to$ increase model size by $1.52\times$ ($2^{0.60}$), compared to Chinchilla's $1.41\times$ ($2^{0.50}$). Over a $10\times$ time increase, this means $3.98\times$ vs.\ $3.16\times$ model size---a 26\% difference that compounds over longer budgets. We release all code and training logs to enable reproduction.

\section{Related Work}
\label{sec:related}

\paragraph{Neural scaling laws.}
\citet{kaplan2020scaling} established power-law scaling of language model loss with model size, dataset size, and compute. Their key finding---that larger models are more sample-efficient---underpins the trend toward ever-larger models. \citet{hoffmann2022training} (Chinchilla) refined this by showing that Kaplan et al.\ underestimated the importance of data: compute-optimal training requires roughly equal scaling of parameters and tokens, yielding $N^* \propto C^{0.50}$ and $D^* \propto C^{0.50}$. Our work extends this framework by replacing compute $C$ with wall-clock time $t$ as the independent variable, finding $N^* \propto t^{0.60}$.

\paragraph{Data-constrained scaling.}
\citet{muennighoff2023scaling} studied the regime where training data is limited and must be repeated, showing that data repetition yields diminishing returns---a finding directly relevant to our setting, where small models exhaust the dataset and cycle through it many times. \citet{hernandez2022scaling} analyzed transfer learning scaling; \citet{clark2022unified} proposed unified scaling laws incorporating model, data, and training steps.

\paragraph{Inference-optimal scaling.}
\citet{sardana2023beyond} studied inference-optimal training, where the optimization target includes inference cost. Their setting resonates with ours: both deviate from pure compute-optimality by incorporating practical constraints. However, their constraint is \emph{inference cost per query}, while ours is \emph{wall-clock training time}.

\paragraph{Efficient training on consumer hardware.}
\citet{dettmers2023qlora} enabled single-GPU fine-tuning via 4-bit quantization, dramatically lowering the barrier to LLM training. The autoresearch framework~\citep{karpathy2025autoresearch} provides small-scale pre-training pipelines for research. \citet{dao2022flashattention} made attention $O(N)$ in memory, enabling longer sequences on limited VRAM. Our work operates in this ecosystem, studying pre-training scaling on consumer GPUs.

\paragraph{Architecture comparisons.}
We compare Dense Transformers against MoE~\citep{jiang2024mixtral,dai2024deepseekmoe}, linear attention variants (RetNet~\citep{sun2023retentive}, GLA~\citep{yang2023gated}, RWKV~\citep{peng2023rwkv}), finding Dense Transformers with Flash Attention dominant under time constraints on consumer GPUs.

\paragraph{Batch size and training dynamics.}
\citet{mccandlish2018empirical} analyzed the critical batch size and its relation to compute--time trade-offs. \citet{li2024surge} studied the ``loss surge'' phenomenon during training. Our work complements these by providing a macro-level time--performance--size relationship.

\section{Experimental Setup}
\label{sec:method}

\subsection{Hardware and Software}

\paragraph{Hardware.} We use a single server with 8$\times$ NVIDIA RTX 4090 GPUs (24GB VRAM each), an AMD EPYC 7542 32-core CPU, and 503GB RAM. Each GPU runs one independent training job---we do \emph{not} use multi-GPU parallelism (data-parallel or model-parallel). This ``embarrassingly parallel'' setup maximizes experimental throughput: 8 configurations run simultaneously. The RTX 4090 represents the highest-end consumer GPU available in 2026, making our results directly relevant to the research community.

\paragraph{Software.} We use the autoresearch framework~\citep{karpathy2025autoresearch} with Flash Attention v2, \texttt{torch.compile} (mode=default, BF16 precision), and cosine learning rate schedule with warmup. All training uses a single-epoch paradigm where the learning rate decays to 10\% of peak over the training budget, with 10\% warmup steps.

\subsection{Architecture Family}

We use decoder-only Transformers with a single ``DEPTH'' parameter $D$ that controls all dimensions:
\begin{equation}
    n\_layers = D, \quad model\_dim = 64D, \quad n\_heads = D, \quad head\_dim = 64
    \label{eq:arch}
\end{equation}

This parameterization ensures constant architectural ratios across scales: the aspect ratio (width/depth) is always 64, and the number of attention heads equals the depth. This eliminates confounding variables from architectural choices and isolates the effect of scale.

Table~\ref{tab:configs} in the Appendix shows all configurations. Key data points: D8 has 50.3M parameters with throughput 428K tok/sec; D14 has 200.9M at 110K tok/sec; D20 has 519.0M at 36K tok/sec; D26 has $\sim$1031M at $\sim$5K tok/sec. This $85\times$ throughput range across our model family is precisely what drives the divergence between time-constrained and compute-constrained scaling.

\subsection{Dataset}

We use FineWeb-Edu~\citep{penedo2024fineweb}, a curated web corpus, tokenized with BPE (32K vocabulary). Our training set contains $\sim$48M tokens. This deliberately small dataset enables studying the \textbf{capacity--data--overfitting interplay}: small models exhaust the data rapidly (D8 cycles through 250+ times at 12h), while large models (D26 at 24h) see fewer than 3 epochs. We report validation bits-per-byte (BPB) throughout.

\paragraph{Why a small dataset?} While larger datasets would avoid overfitting, our small dataset serves a crucial purpose: it creates a natural laboratory for studying \emph{both} forms of suboptimality simultaneously---undertrained large models and overfit small models---within practical time budgets. With a sufficiently large dataset, the U-curve's left limb (overfitting) would not appear until much longer training, making the phenomenon harder to observe.

\subsection{Experimental Design}

We train models across 8 time budgets: 5 minutes, 30 minutes, 1 hour, 2 hours, 4 hours, 8 hours, 12 hours, and 24 hours. At each budget, we run 4--8 model sizes in parallel across GPUs. All configurations share identical hyperparameters (learning rate $3 \times 10^{-4}$, batch size 64 sequences of 512 tokens) except DEPTH and cases where large models require reduced batch size for VRAM constraints (D26 uses batch size 1). Learning rate ablations (Section~\ref{sec:ablation}) confirm robustness to this choice.

In total, we conduct 70+ individual training runs, spanning a $20\times$ parameter range (50M--1031M) and a $288\times$ time range (5min--24h).

\section{Results}
\label{sec:results}

\subsection{U-Shaped Optimality at Every Time Budget}

Our central empirical finding is that at every time budget, validation BPB versus model size yields a characteristic U-shaped curve. Table~\ref{tab:main_results} presents complete results across all 8 time budgets and 10 model sizes.

\begin{table}[t]
\centering
\caption{Validation BPB across model sizes and time budgets. \textbf{Bold} = optimal at each budget. $\uparrow$ = overfitting (worse than shorter training). At 12h, no U-shape: BPB decreases monotonically with model size. At 24h, U-shape returns due to data bottleneck.}
\label{tab:main_results}
\small
\begin{tabular}{lccccccccc}
\toprule
Model & Params & 5min & 30min & 1h & 2h & 4h & 8h & 12h & 24h \\
\midrule
D8  & 50M  & \textbf{1.133} & 0.977 & 0.979$\uparrow$ & 0.906 & 0.925$\uparrow$ & 0.886 & 0.919$\uparrow$ & --- \\
D10 & 86M  & 1.178 & \textbf{0.973} & 0.976 & 0.906 & 0.892 & 0.886 & 0.885 & --- \\
D12 & 135M & 1.363 & 1.001 & 0.991 & 0.904 & 0.878 & 0.873 & 0.871 & 0.870 \\
D14 & 201M & 1.578 & 1.016 & \textbf{0.945} & \textbf{0.901} & 0.866 & 0.854 & 0.852 & 0.857$\uparrow$ \\
D16 & 285M & 1.566 & 1.026 & 0.951 & \textbf{0.901} & \textbf{0.862} & 0.844 & 0.841 & 0.851$\uparrow$ \\
D18 & 384M & --- & --- & --- & --- & 0.866 & 0.837 & 0.833 & 0.845$\uparrow$ \\
D20 & 519M & 1.804 & --- & 1.009 & --- & 0.872 & \textbf{0.836} & 0.828 & 0.838$\uparrow$ \\
D22 & 621M & --- & --- & --- & --- & --- & --- & 0.826 & 0.829 \\
D24 & 856M & 1.854 & --- & --- & --- & 0.896 & 0.845 & \textbf{0.824} & 0.817 \\
D26 & 1031M & --- & --- & --- & --- & --- & --- & --- & \textbf{0.814} \\
\bottomrule
\end{tabular}
\end{table}

The U-shape arises from two competing forces. The \emph{left limb} (too-small models): these models exhaust the training data and overfit. D8 at 4h achieves BPB 0.925, \emph{worse} than its 2h result of 0.906; by 12h, its BPB degrades further to 0.919 after cycling through the 48M-token dataset over 250 times. The \emph{right limb} (too-large models): insufficient throughput means too little data is processed to utilize model capacity---D24 at 5min sees only 13M tokens, yielding BPB 1.854, far worse than D8's 1.133 with 134M tokens.

The optimal model shifts right monotonically with time: D8 (5min) $\to$ D10 (30min) $\to$ D14 (1h) $\to$ D14=D16 (2h) $\to$ D16 (4h) $\to$ D20 (8h) $\to$ D24 (12h) $\to$ D26 (24h). This monotonic rightward shift is the key empirical regularity that enables fitting a scaling law.

\subsection{Time-Constrained Scaling Law}
\label{sec:scaling_law}

Fitting a power law $N^*(t) = a \cdot t^{\alpha}$ to all 8 time points (5min--24h):
\begin{equation}
    N^*(t) = 14.20 \times t^{0.595} \quad (R^2 = 0.963)
    \label{eq:scaling_params}
\end{equation}
where $N^*$ is optimal parameters (millions) and $t$ is time (minutes). The exponent $\alpha = 0.595 \pm 0.067$ \textbf{exceeds Chinchilla's compute-optimal exponent of 0.50} by approximately 20\%.

\paragraph{Statistical significance.} The 95\% confidence interval for $\alpha$ is $[0.53, 0.67]$, which excludes the Chinchilla value of 0.50. This confirms that time-constrained scaling represents a genuinely different regime from compute-constrained scaling.

\paragraph{Practical interpretation.} Doubling the time budget prescribes $2^{0.60} = 1.52\times$ the model size, compared to Chinchilla's $2^{0.50} = 1.41\times$. Over a $10\times$ increase, this means $3.98\times$ vs.\ $3.16\times$ model size. The difference compounds: for a researcher moving from a 1-hour ``quick experiment'' to a 24-hour overnight run, our law recommends a $24^{0.60}/1^{0.60} = 7.2\times$ larger model, vs.\ Chinchilla's $24^{0.50} = 4.9\times$.

\paragraph{Depth scaling.} The optimal DEPTH also follows a power law: $D^*(t) = 4.97 \times t^{0.231}$ ($R^2 = 0.958$).

\paragraph{Sensitivity analysis.} We verify robustness by varying fitting choices (full results in Table~\ref{tab:sensitivity}, Appendix). At 2h, D14 and D16 tie at BPB 0.901; we test using average (243M), D16 only, D14 only, and excluding the 2h point. Across all variants, $\alpha$ robustly exceeds 0.50 ($\alpha \in [0.60, 0.81]$). Notably, $\alpha$ \emph{decreased} from 0.747 (7 points) to 0.595 (8 points) upon adding 24h data. We discuss this convergence behavior in Section~\ref{sec:discussion}.

\subsection{Loss Scaling}

The best achievable loss at each time budget follows:
\begin{equation}
    L^*(t) = 1.223 \times t^{-0.061} \quad (R^2 = 0.971)
    \label{eq:scaling_loss}
\end{equation}

Diminishing returns are pronounced (full breakdown in Table~\ref{tab:marginal}, Appendix): the first 30 minutes yield 0.16 bpb improvement (0.38 bpb/hour), while hours 8--24 yield only 0.022 bpb combined (0.001 bpb/hour). On consumer GPUs with limited data, the first hour captures the majority of achievable improvement---quick exploratory runs (1--4h) provide most of the signal.

\subsection{The Dual U-Shape: Two Bottlenecks, One Curve}
\label{sec:dual_ushape}

A striking finding from our 24h experiments is the emergence of a \emph{second} U-shape driven by a completely different mechanism than the first.

\paragraph{Short-budget U-shape (5min--8h): compute bottleneck.} At budgets $\leq$8h, the right limb of the U-curve arises because large models have insufficient throughput to process enough data. D24 (856M) at 4h achieves BPB 0.896---far from its potential---because it processes only $\sim$13M tokens.

\paragraph{U-curve disappearance ($\sim$12h): transitional regime.} At 12h, a remarkable phenomenon occurs: the U-curve \emph{disappears}. BPB decreases monotonically from D10 (0.885) to D24 (0.824). The largest model shows the most dramatic improvement ($\Delta_{8h \to 12h} = -0.021$), suggesting the true optimum exceeds our VRAM capacity.

\paragraph{Long-budget U-shape (24h): data bottleneck.} At 24h, the U-curve \emph{returns}---but its left limb has a new cause. Medium models (D14--D20), which were optimal at shorter budgets, now \emph{overfit}: D14's BPB degrades from 0.852 (12h) to 0.857 (24h); D18 degrades from 0.833 to 0.845 (+0.012). Only D24 and D26 continue improving (see Table~\ref{tab:overfitting} in Appendix for full breakdown). The U-curve's left limb at 24h is driven by \emph{data repetition}, not insufficient compute.

This dual U-shape mechanism means the ``safe zone'' of near-optimal model sizes \emph{narrows} at long budgets: at 4h, D14--D18 all achieve within 0.004 of optimal; at 24h, only D24--D26 avoid overfitting.

\subsection{Ablation Studies}
\label{sec:ablation}

\paragraph{Learning rate sensitivity.} We conduct LR ablations at two model sizes (D8 and D14) with 1h training, testing LR values from $1.5 \times 10^{-4}$ to $6 \times 10^{-4}$. D14's default LR ($3 \times 10^{-4}$) is near-optimal, within 0.001 BPB of the best LR ($6 \times 10^{-4}$, BPB 0.944 vs.\ 0.945). Even with optimal LR, D8 (BPB 0.948) does not beat D14 (0.945), confirming that the scaling law is robust to moderate LR variation.

\paragraph{Multi-seed validation.} We run 3 seeds (42, 123, 456) for four model sizes (D8, D10, D14, D16) at 30 minutes (Table~\ref{tab:multiseed}). The results are remarkably stable: inter-seed standard deviation is at most 0.0032 BPB, while inter-model gaps reach 0.056 BPB---a factor of $17\times$ larger. The coefficient of variation is below 0.31\% for all models. Crucially, D10 remains the optimal model at 30min across all seeds ($\text{mean} = 0.9734 \pm 0.0004$), and every D10 seed beats every D8 seed, confirming the Chinchilla crossover is robust to randomness. These results validate that our primary single-seed results are representative of the underlying trends.

\begin{table}[t]
\centering
\caption{Multi-seed validation at 30 minutes. Three seeds per model. Max $\sigma = 0.003$ BPB vs.\ inter-model gaps of 0.002--0.056 BPB. D10 is optimal across all seeds.}
\label{tab:multiseed}
\small
\begin{tabular}{lcccccc}
\toprule
Model & Params & Seed 42 & Seed 123 & Seed 456 & Mean $\pm$ Std & CV (\%) \\
\midrule
D8 & 50M & 0.977 & 0.975 & 0.975 & $0.976 \pm 0.001$ & 0.13 \\
\textbf{D10} & \textbf{86M} & \textbf{0.973} & \textbf{0.974} & \textbf{0.973} & $\mathbf{0.973 \pm 0.000}$ & \textbf{0.04} \\
D14 & 201M & 1.016 & 1.022 & 1.017 & $1.018 \pm 0.003$ & 0.29 \\
D16 & 285M & 1.026 & 1.032 & 1.029 & $1.029 \pm 0.003$ & 0.31 \\
\bottomrule
\end{tabular}
\end{table}

\paragraph{Multi-architecture comparison.} At 5min on RTX 4090, we compare Dense Transformer against four alternatives: MoE (64 experts, top-1, fine-grained), RetNet with FLA Triton kernels, GLA with FLA kernels, and RWKV6 with FLA kernels. Results: Dense Transformer (1.133) $\gg$ MoE (1.143) $\gg$ RetNet (2.216) $>$ GLA (2.249) $>$ RWKV6 (2.258). Linear attention variants underperform by nearly $2\times$ despite achieving higher model FLOPs utilization (MFU: 20--27\% vs.\ 10\% for Dense). The architectural comparison justifies our focus on Dense Transformers.

\section{Discussion}
\label{sec:discussion}

\subsection{Why Time $\neq$ Compute}

The mathematical origin of the time--compute divergence is straightforward. For a model of size $N$ trained for time $t$, the compute consumed is:
\begin{equation}
    C(N, t) = 6N \times \tau(N) \times t
    \label{eq:compute}
\end{equation}
where $\tau(N)$ is throughput (tokens/second) and the factor 6 accounts for forward and backward passes. We empirically measure $\tau(N) \propto N^{-0.8}$ on our hardware (Table~\ref{tab:configs}), giving:
\begin{equation}
    C \propto N \times N^{-0.8} \times t = N^{0.2} \times t
    \label{eq:compute_scaling}
\end{equation}

Under \emph{compute constraints} (fixed $C$), doubling $N$ halves the available tokens: $T = C / 6N$. The penalty for increasing model size is direct and linear. Under \emph{time constraints} (fixed $t$), doubling $N$ reduces tokens by only $2^{0.8} \approx 1.74\times$: $T = \tau(N) \times t$. The penalty is indirect, mediated through throughput, and \emph{sublinear}. This weaker penalty means time constraints favor larger models than compute constraints---precisely what $\alpha = 0.60 > 0.50$ captures.

\subsection{Comparison with Chinchilla}

The quantitative implications of $\alpha = 0.60$ vs.\ 0.50 compound with longer budgets. Doubling time prescribes $1.52\times$ the model (vs.\ Chinchilla's $1.41\times$); over a $10\times$ increase: $3.98\times$ vs.\ $3.16\times$; for the common scenario of scaling from 1h to 24h: $7.2\times$ vs.\ $4.9\times$ (Table~\ref{tab:comparison} in Appendix).

The mechanism behind this divergence is the \emph{data repetition penalty}. Under time constraints, small models process the same data many more times, suffering compounding repetition losses~\citep{muennighoff2023scaling}. This creates a ``push'' toward larger models that is absent in compute-constrained settings with unlimited data.

\subsection{Evolution of $\alpha$}

An intriguing phenomenon is the non-monotonic convergence of $\alpha$ as we add longer time points:

\begin{itemize}
    \item 5 points (5min--4h): $\alpha = 0.44$ --- below Chinchilla
    \item 6 points (+8h): $\alpha = 0.55$ --- just above Chinchilla
    \item 7 points (+12h): $\alpha = 0.75$ --- far above Chinchilla
    \item 8 points (+24h): $\alpha = 0.60$ --- settled above Chinchilla
\end{itemize}

The 7-point estimate ($\alpha = 0.75$) was inflated because 12h was the first budget where the U-curve disappeared and D24 (the largest model) became optimal---this created an outsized leverage point. Adding 24h data with D26 (1031M) as the new optimum ``pulled back'' $\alpha$ to a more moderate 0.60. We believe $\alpha \approx 0.60$ is a reasonable estimate, though it may continue to evolve with even longer budgets.

The fact that $\alpha$ oscillates rather than converges monotonically suggests the relationship between $N^*$ and $t$ may not be a pure power law over all time scales. The dual U-shape mechanism (Section~\ref{sec:dual_ushape}) introduces regime transitions that a single power law cannot perfectly capture. Nevertheless, $R^2 = 0.963$ indicates an excellent fit over our observed range.

\subsection{The Disappearing and Returning U-Curve}

The lifecycle of the U-curve across time budgets reveals three distinct regimes:

\textbf{Regime 1: Compute-bounded (5min--8h).} U-curves with clear minima. The right limb arises from throughput limitations. Optimal model sizes are well below VRAM capacity.

\textbf{Regime 2: Transitional ($\sim$12h).} The U-curve vanishes: BPB decreases monotonically with model size. This represents the ``sweet spot'' where even the largest fittable model (D24, 856M) benefits from additional training without overfitting. The true optimum likely exceeds single-GPU VRAM---a natural transition point to multi-GPU training.

\textbf{Regime 3: Data-bounded ($\geq$24h).} The U-curve returns, but with inverted causation: the left limb is now driven by overfitting (too many epochs on finite data), not insufficient compute. Models D14--D20 degrade from their 12h BPBs. This regime is specific to finite datasets and would shift rightward with larger data.

This three-regime structure has a practical implication: there exists a ``golden duration'' for each dataset size beyond which increasing training time requires either more data or model scale beyond single-GPU capacity.

\subsection{Practical Guidelines}

\textbf{Rule of thumb:} double your time $\to$ $1.52\times$ model size ($2^{0.60}$). For exploratory work, 1--4 hours captures 80\%+ of achievable improvement. For overnight runs (12--24h), maximize model size within VRAM constraints. Table~\ref{tab:guidelines} (Appendix) provides specific recommendations per time budget for RTX 4090.

\subsection{Limitations}

Our study has several limitations that future work should address:

\begin{enumerate}
    \item \textbf{Single GPU type.} RTX 4090 has specific throughput--size scaling. A100/H100 GPUs may yield different $\alpha$ due to different memory bandwidth, compute ratios, and multi-GPU interconnects.
    
    \item \textbf{Small dataset.} Our 48M-token dataset amplifies data repetition effects. With larger datasets (billions of tokens), the data-bounded regime (Section~\ref{sec:dual_ushape}) would shift rightward, potentially reducing $\alpha$ toward 0.50.
    
    \item \textbf{Single architecture.} While we confirm Dense Transformers dominate alternatives on consumer GPUs, the specific $\alpha$ value may differ for other architectures.
    
    \item \textbf{Limited multi-seed coverage.} While our multi-seed experiment at 30 minutes confirms negligible inter-seed variance ($\sigma < 0.003$), full multi-seed validation at all 8 time budgets would further strengthen confidence intervals, particularly at the longer budgets (12--24h) where single runs are expensive.
    
    \item \textbf{No multi-GPU training.} At 12--24h, the optimal model approaches or exceeds single-GPU VRAM, suggesting multi-GPU training is necessary for longer budgets. The scaling law may behave differently in the multi-GPU regime due to communication overhead.
\end{enumerate}

\section{Conclusion}
\label{sec:conclusion}

We introduce \textbf{time-constrained scaling laws} for language model training on consumer GPUs. Our central finding is that optimal model size scales as $N^*(t) \propto t^{0.60}$, exceeding Chinchilla's compute-optimal exponent of 0.50. This divergence arises from sublinear throughput scaling ($\tau \propto N^{-0.8}$), which breaks the equivalence between time and compute as optimization targets.

Three key phenomena emerge from our 70+ training runs:

\begin{enumerate}
    \item \textbf{U-shaped optimality curves} at every time budget, with the optimal model size monotonically shifting rightward from 50M (5min) to 1031M (24h).
    
    \item \textbf{A dual U-shape mechanism}: short-budget U-curves from compute bottlenecks, long-budget U-curves from data bottlenecks, with a transitional regime where the U-curve disappears.
    
    \item \textbf{Severe diminishing returns}: the first hour captures most achievable improvement; extending from 8h to 24h yields only 0.022 BPB---a finding that should inform how researchers allocate GPU time.
\end{enumerate}

Most practitioners operate under time constraints, not compute budgets. Guidance in wall-clock hours---not teraFLOPs---makes scaling laws more actionable for the research community.

\paragraph{Broader impact.} Our findings suggest that the ``bigger is better'' narrative, while correct directionally, overstates the benefit of model scale under time constraints. For the growing community of researchers training on consumer hardware, understanding time-constrained scaling can save thousands of GPU-hours of misallocated compute.

\paragraph{Future work.} Natural extensions include: (1) multi-GPU training at 12h+ budgets, where the optimal model exceeds single-GPU VRAM; (2) datacenter GPUs (A100/H100) with different throughput profiles; (3) larger datasets to isolate throughput effects from data repetition; (4) testing whether $\alpha$ converges to a stable value with more time points; and (5) extending to fine-tuning and RLHF stages of the LLM pipeline.


\begin{figure}[t]
    \centering
    \includegraphics[width=0.95\textwidth]{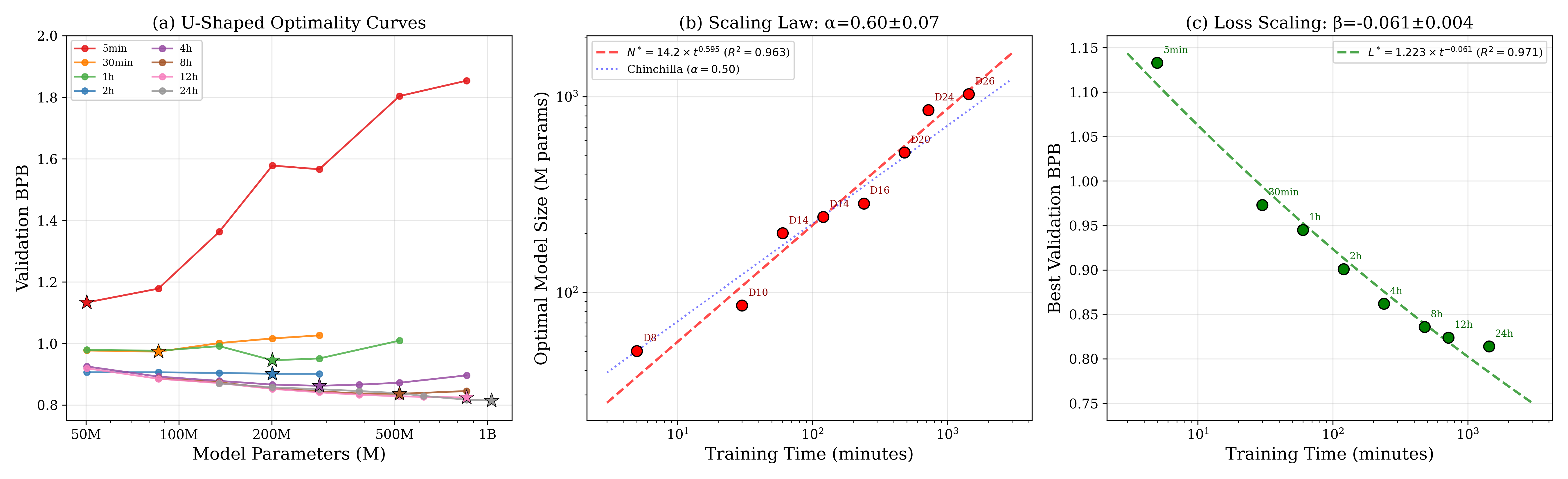}
    \caption{\textbf{Main result.} \textbf{(a)}~U-shaped optimality curves at 8 time budgets (5min--24h). Stars mark optimal model at each budget. \textbf{(b)}~Optimal model size vs.\ time, fitted with $N^* = 14.2 \times t^{0.595}$ ($R^2 = 0.963$). The blue dotted line shows Chinchilla's $\alpha = 0.50$ for reference. \textbf{(c)}~Best achievable BPB vs.\ time, fitted with $L^* = 1.22 \times t^{-0.061}$ ($R^2 = 0.971$), showing severe diminishing returns.}
    \label{fig:composite}
\end{figure}

\begin{figure}[t]
    \centering
    \begin{subfigure}[t]{0.48\textwidth}
        \centering
        \includegraphics[width=\textwidth]{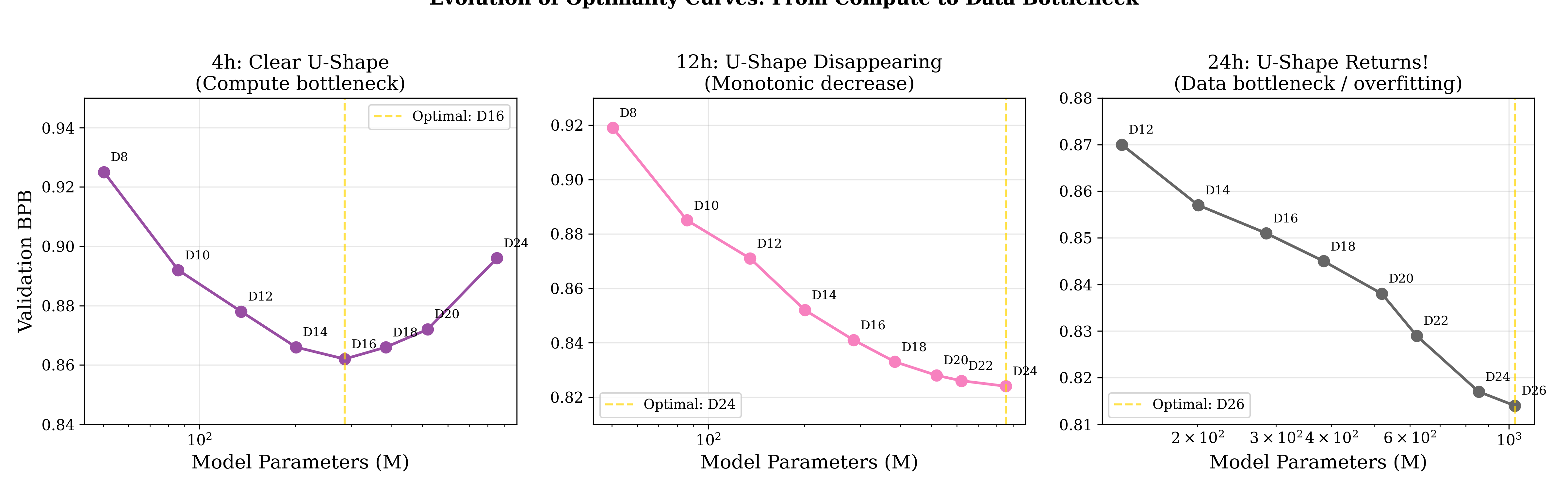}
        \caption{U-curve evolution: clear U-shape at 4h (compute bottleneck), monotonic decrease at 12h (transitional), U-shape returns at 24h (data bottleneck).}
        \label{fig:ucurve_evolution}
    \end{subfigure}
    \hfill
    \begin{subfigure}[t]{0.48\textwidth}
        \centering
        \includegraphics[width=\textwidth]{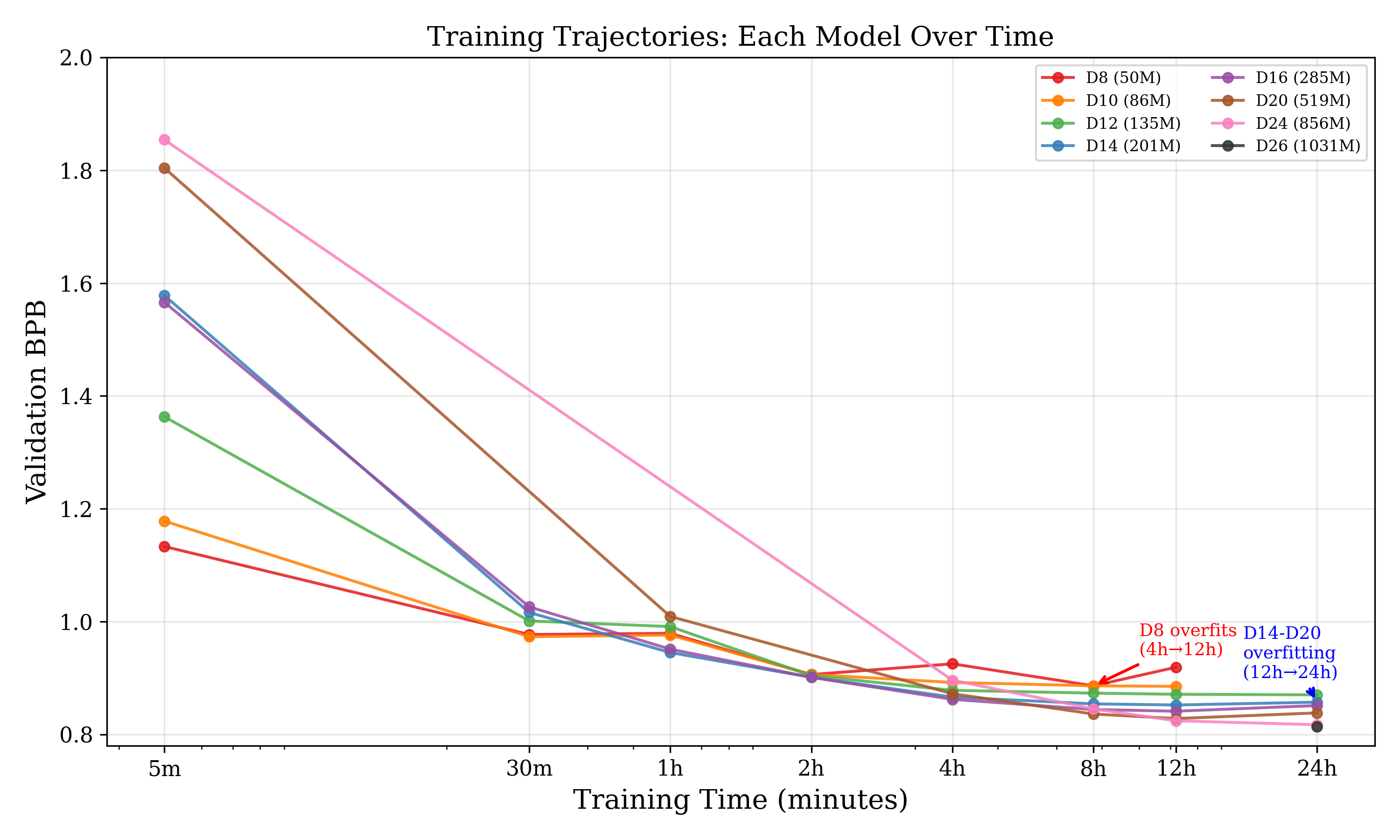}
        \caption{Training trajectories. D8 overfits after 2h; D14--D20 overfit between 12h and 24h; D24 and D26 continue improving at 24h.}
        \label{fig:trajectories}
    \end{subfigure}
    \caption{\textbf{Key phenomena.} Left: the dual U-shape mechanism across three regimes. Right: training trajectories showing model-specific overfitting dynamics. Additional figures ($\alpha$ convergence, heatmap, dual U-shape detail) in Appendix.}
    \label{fig:phenomena}
\end{figure}

\bibliographystyle{plainnat}
\bibliography{references}

\appendix
\section{Model Configurations}
\label{app:configs}

\begin{table}[h]
\centering
\caption{Full model configurations parameterized by DEPTH. Throughput measured on single RTX 4090.}
\label{tab:configs}
\small
\begin{tabular}{cccccc}
\toprule
DEPTH & Layers & Dim & Params (M) & tok/sec & MFU (\%) \\
\midrule
8  & 8  & 512  & 50.3  & 428K & 10.4 \\
10 & 10 & 640  & 85.9  & 252K & 10.7 \\
12 & 12 & 768  & 135.3 & 160K & 10.9 \\
14 & 14 & 896  & 200.9 & 110K & 11.4 \\
16 & 16 & 1024 & 285.2 & 78K  & 11.4 \\
18 & 18 & 1152 & $\sim$384 & 56K & $\sim$11 \\
20 & 20 & 1280 & 519.0 & 36K  & $\sim$11 \\
22 & 22 & 1408 & $\sim$621 & 27K & $\sim$11 \\
24 & 24 & 1536 & 855.6 & 20K  & $\sim$11 \\
26 & 26 & 1664 & $\sim$1031 & $\sim$5K & $\sim$3 \\
\bottomrule
\end{tabular}
\end{table}

\section{Sensitivity Analysis}
\label{app:sensitivity}

\begin{table}[h]
\centering
\caption{Sensitivity of the scaling exponent $\alpha$ to fitting choices. All variants yield $\alpha > 0.50$.}
\label{tab:sensitivity}
\small
\begin{tabular}{lccc}
\toprule
Variant & $\alpha$ & 95\% CI & $R^2$ \\
\midrule
8-point (baseline) & $0.595 \pm 0.067$ & $[0.53, 0.67]$ & 0.963 \\
7-point (exclude 24h) & $0.747 \pm 0.107$ & $[0.64, 0.86]$ & 0.957 \\
7-point (2h = 285M) & $0.706 \pm 0.110$ & $[0.60, 0.82]$ & 0.948 \\
6-point (exclude 2h) & $0.805 \pm 0.145$ & $[0.66, 0.95]$ & 0.960 \\
\bottomrule
\end{tabular}
\end{table}

\section{Marginal Returns}
\label{app:marginal}

\begin{table}[h]
\centering
\caption{Marginal BPB improvement per time increment. Returns diminish rapidly after 1 hour.}
\label{tab:marginal}
\small
\begin{tabular}{lcccc}
\toprule
Interval & $\Delta t$ & $\Delta$ BPB & BPB/hour & Efficiency \\
\midrule
5min $\to$ 30min & 25min & $-0.160$ & $-0.384$ & Very high \\
30min $\to$ 1h & 30min & $-0.028$ & $-0.056$ & High \\
1h $\to$ 2h & 1h & $-0.044$ & $-0.044$ & Moderate \\
2h $\to$ 4h & 2h & $-0.039$ & $-0.020$ & Low \\
4h $\to$ 8h & 4h & $-0.026$ & $-0.007$ & Very low \\
8h $\to$ 12h & 4h & $-0.012$ & $-0.003$ & Diminishing \\
12h $\to$ 24h & 12h & $-0.010$ & $-0.001$ & Minimal \\
\bottomrule
\end{tabular}
\end{table}

\section{Overfitting Analysis (12h $\to$ 24h)}
\label{app:overfitting}

\begin{table}[h]
\centering
\caption{Change in BPB from 12h to 24h. Positive values indicate overfitting. Only models $\geq$856M continue improving.}
\label{tab:overfitting}
\small
\begin{tabular}{lcccc}
\toprule
Model & Params & BPB$_{12h}$ & BPB$_{24h}$ & $\Delta$ \\
\midrule
D12 & 135M & 0.871 & 0.870 & $-0.001$ (saturated) \\
D14 & 201M & 0.852 & 0.857 & $+0.005$ (overfitting) \\
D16 & 285M & 0.841 & 0.851 & $+0.010$ (overfitting) \\
D18 & 384M & 0.833 & 0.845 & $+0.012$ (overfitting) \\
D20 & 519M & 0.828 & 0.838 & $+0.010$ (overfitting) \\
D22 & 621M & 0.826 & 0.829 & $+0.003$ (marginal) \\
D24 & 856M & 0.824 & 0.817 & $-0.007$ (improving) \\
D26 & 1031M & --- & 0.814 & first run \\
\bottomrule
\end{tabular}
\end{table}

\section{Comparison with Chinchilla}
\label{app:comparison}

\begin{table}[h]
\centering
\caption{Model size recommendations: time-constrained (ours, $\alpha=0.60$) vs.\ compute-constrained (Chinchilla, $\alpha=0.50$).}
\label{tab:comparison}
\small
\begin{tabular}{lcc}
\toprule
Time multiplier & Ours ($2^{0.60}$) & Chinchilla ($2^{0.50}$) \\
\midrule
$2\times$ time & $1.52\times$ model & $1.41\times$ model \\
$4\times$ time & $2.30\times$ model & $2.00\times$ model \\
$10\times$ time & $3.98\times$ model & $3.16\times$ model \\
$24\times$ time (1h$\to$24h) & $7.22\times$ model & $4.90\times$ model \\
\bottomrule
\end{tabular}
\end{table}

\section{Additional Figures}
\label{app:figures}

\begin{figure}[h]
    \centering
    \includegraphics[width=0.7\textwidth]{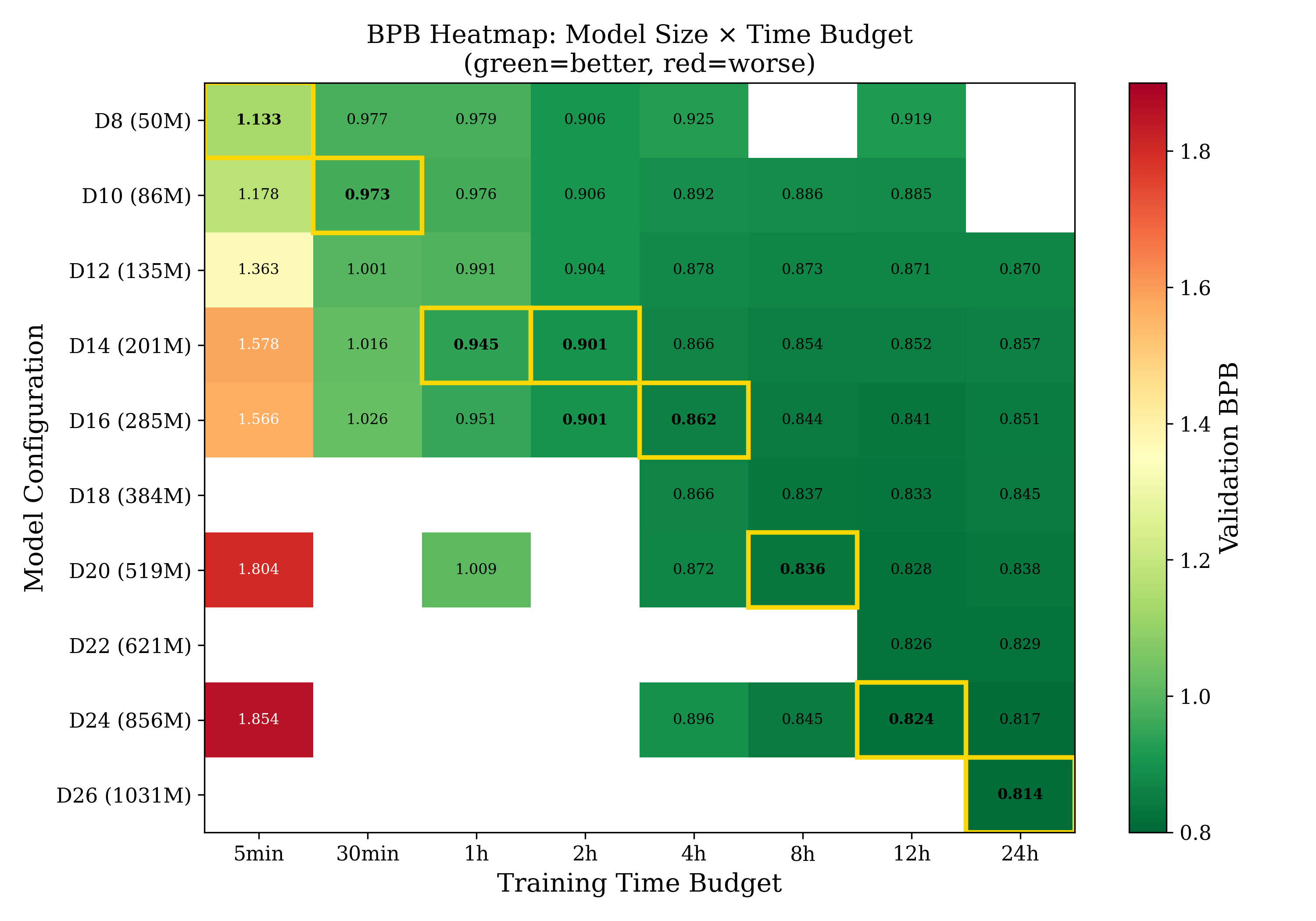}
    \caption{BPB heatmap across model sizes and time budgets. The diagonal band of optimal values traces the time-constrained scaling law. Gold boxes mark optimal configurations.}
    \label{fig:heatmap}
\end{figure}

\begin{figure}[h]
    \centering
    \includegraphics[width=0.6\textwidth]{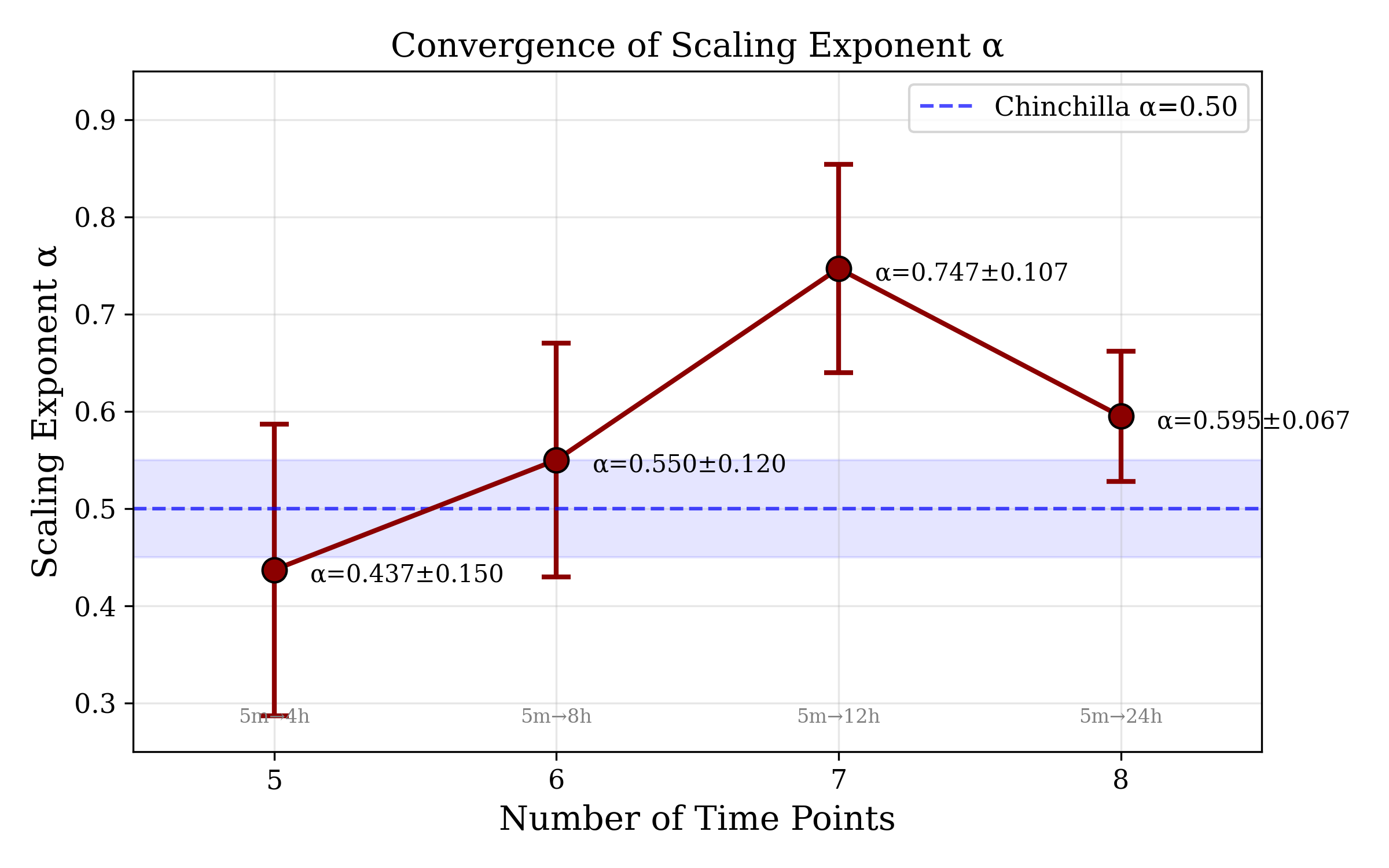}
    \caption{$\alpha$ evolution as time points are added: 0.44 (5pt) $\to$ 0.55 (6pt) $\to$ 0.75 (7pt) $\to$ 0.60 (8pt). The non-monotonic convergence reflects regime transitions (Section~\ref{sec:discussion}).}
    \label{fig:alpha_convergence}
\end{figure}

\begin{figure}[h]
    \centering
    \includegraphics[width=0.95\textwidth]{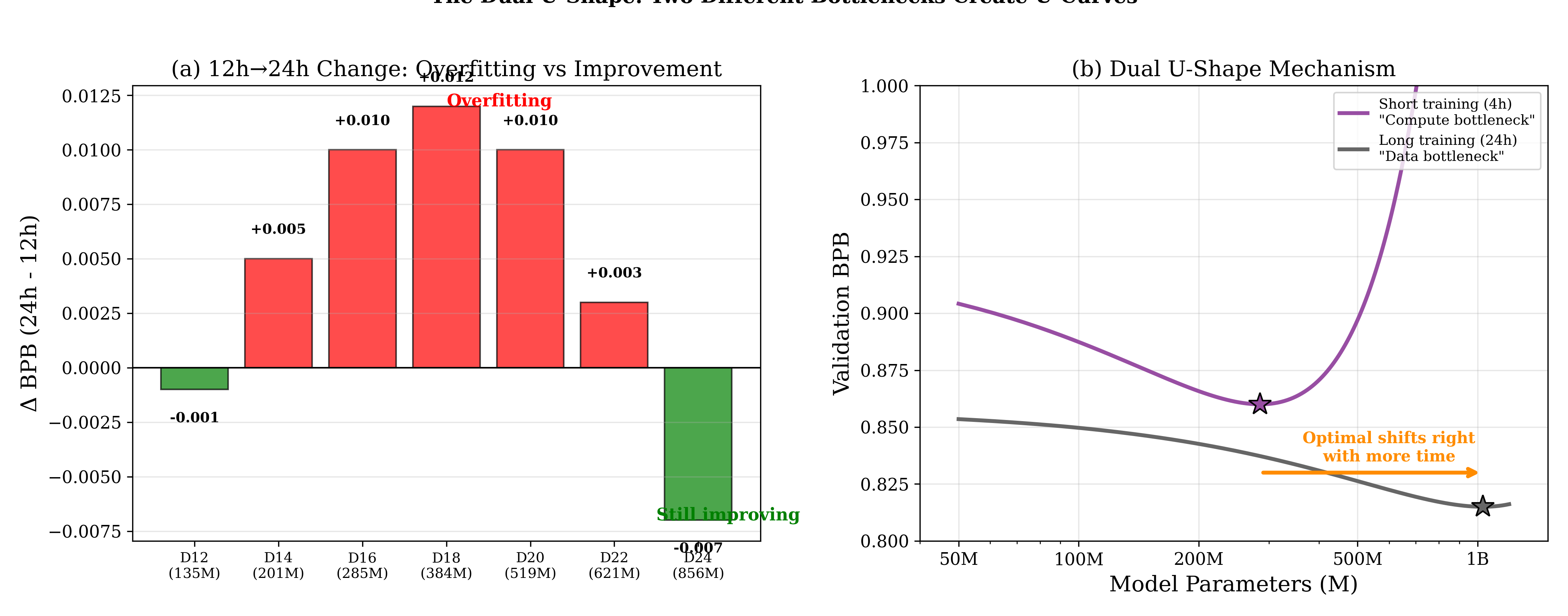}
    \caption{\textbf{Dual U-shape mechanism.} \textbf{(a)}~Change in BPB from 12h to 24h per model. Models D14--D20 overfit (red bars, positive $\Delta$), while D24--D26 continue improving (green bars, negative $\Delta$). \textbf{(b)}~Conceptual illustration of the dual U-shape.}
    \label{fig:dual_ushape}
\end{figure}

\section{Multi-Seed Validation Details}
\label{app:multiseed}

\begin{figure}[h]
    \centering
    \includegraphics[width=0.95\textwidth]{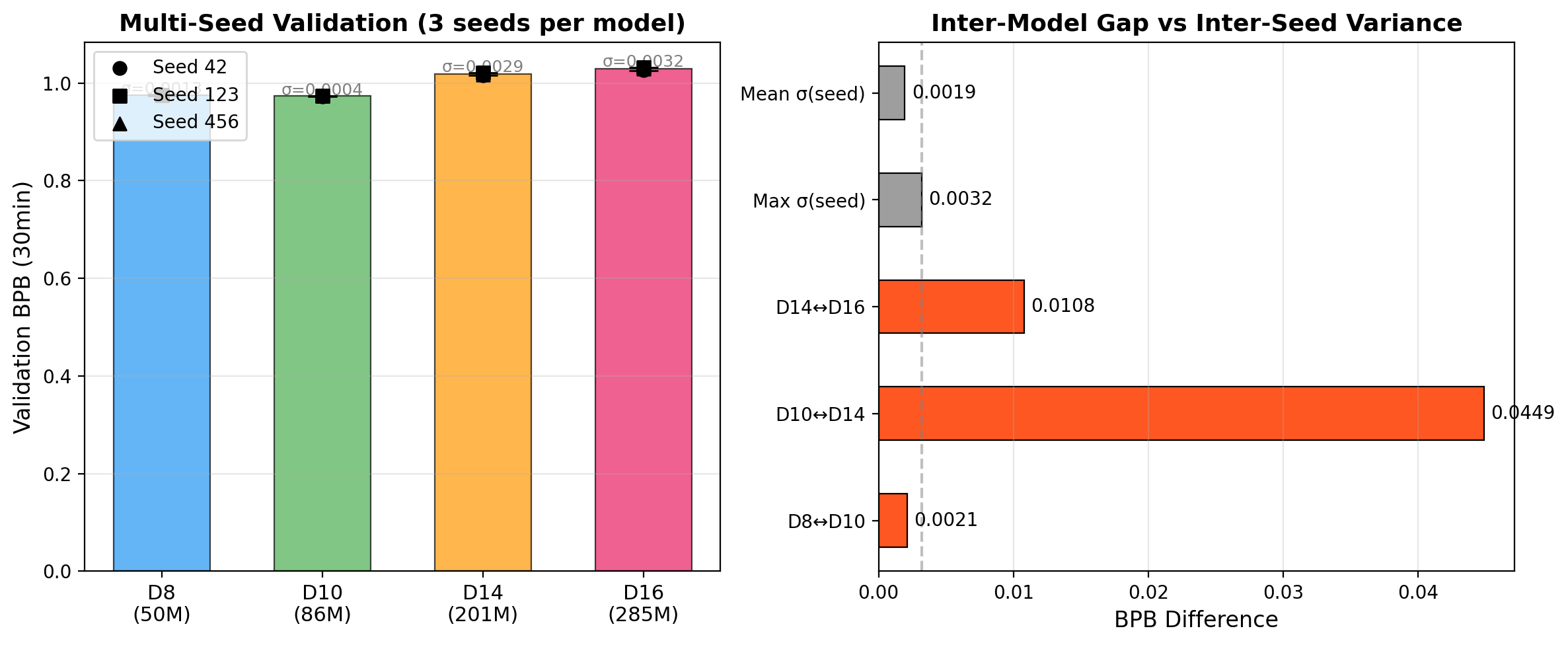}
    \caption{\textbf{Multi-seed validation.} \textbf{(a)}~Validation BPB for 3 seeds per model at 30 minutes. Error bars show $\pm 1\sigma$. Individual seed values (markers) overlap almost perfectly. \textbf{(b)}~Inter-model gaps (red) vs.\ inter-seed variance (gray). Model choice matters orders of magnitude more than random seed.}
    \label{fig:multiseed}
\end{figure}

The multi-seed experiment serves as a critical robustness check. At the 30-minute budget---where D10 (86M) and D8 (50M) are separated by only 0.002 BPB---all three D10 seeds produce lower BPB than all three D8 seeds. This pairwise dominance across $3 \times 3 = 9$ comparisons gives strong evidence that the Chinchilla crossover at 30 minutes is a genuine phenomenon, not a statistical artifact.

The exceptionally low variance of D10 ($\sigma = 0.0004$) compared to D14 and D16 ($\sigma \approx 0.003$) likely reflects that D10 at 30min is operating at a natural ``sweet spot'' where it has processed sufficient data ($\sim$450M tokens, $\sim$5$\times$ the dataset) to produce stable representations, whereas D14 and D16 have seen fewer tokens relative to their capacity and are thus more sensitive to initialization.

\section{Practical Guidelines}
\label{app:guidelines}

\begin{table}[h]
\centering
\caption{Recommended model sizes for RTX 4090 training with FineWeb-Edu-scale data.}
\label{tab:guidelines}
\small
\begin{tabular}{lccc}
\toprule
Time budget & Model size & DEPTH & Expected BPB \\
\midrule
5 minutes & 50M & D8 & 1.133 \\
30 minutes & 86M & D10 & 0.973 \\
1 hour & 200M & D14 & 0.945 \\
2 hours & 200--285M & D14--D16 & 0.901 \\
4 hours & 285M & D16 & 0.862 \\
8 hours & 519M & D20 & 0.836 \\
12 hours & 856M & D24 & 0.824 \\
24 hours & $\sim$1B & D26 & 0.814 \\
\bottomrule
\end{tabular}
\end{table}

\end{document}